\documentstyle[12pt]{article}
\begin{document}
\title
{\bf
Effective Chiral-Spin Hamiltonian for Odd-numbered Coupled Heisenberg Chains}
\author{
V. Subrahmanyam\cr
International Center for Theoretical Physics, P.O. Box 586\cr
34100 Trieste, Italy\cr}
\maketitle
%PACS No : 74.70, 75.30.D\\
\begin{abstract}
An $L \times \infty$  system of odd number of coupled Heisenberg spin chains is
studied using a degenerate perturbation theory, where $L$ is the number of
coupled chains. An effective chain Hamiltonian is derived explicitly in terms
of two spin half degrees of freedom of a closed chain of $L$ sites, valid in
the regime the inter-chain coupling is stronger than the intra-chain
coupling. The spin gap has been calculated numerically using the effective
Hamiltonian for $L=3,5,7,9$ for a finite chain up to ten sites. It is suggested
that the ground state of the effective Hamiltonian is correlated, by examining
various variational trial states for the effective-spin chain Hamiltonian.

\end{abstract}
\vskip 1in

\newpage
%\textwidth = 14truecm
\hoffset = -1truecm
%%-main body of paper-%%
\parindent=0.5in
There is a growing interest in coupled chains of Hubbard-Heisenberg spin
systems, following the experimental realization of coupled arrays of
metal-oxide-metal ladder systems \cite{kn:Joh87,kn:Tak92}. A number of
investigations \cite{kn:Dag92,kn:Str92,kn:Bar93,kn:Ric93,kn:Par93} of
weakly-coupled Heisenberg spin chains have been carried out,  in which the
coupling between the chains is weaker than intra-chain coupling, and have
provided a strong
indication that a system of even number of chains can be understood in terms of
a short-range resonating valence bond (RVB) picture, with spin gap and a finite
spin-spin correlation length. Whereas a system of odd number of chains is
gapless
with power-law spin correlations, indicating that it is in the same
universality
class as a single spin chain. The framework for understanding the difference
between even and odd numbered chains is the RVB
picture\cite{kn:Ric93,kn:Whi94},
which suggests
that an even-numbered coupled chain system can be thought of as an integer spin
chain exhibiting a Haldane spin gap, and an odd-numbered coupled system maps
onto a half-odd integer spin chain.
Below we will construct an effective spin Hamiltonian for odd-numbered coupled
system, which is cast explcitily in terms of spin half degrees of freedom. This
can be done straightforwardly in
the regime the coupling along the chain is weaker than the coupling between the
chains, complementary to the earlier studies.

We consider the Heisenberg Hamiltonian given as
\begin{equation}
H=J\sum_{l,n}{\vec s}_{l,n} {\vec s}_{l+1,n}  +
J^\prime\sum_{l,n}{\vec s}_{l,n} {\vec s}_{l,n+1} -h\sum_{l,n}s_{l,n}^z
\label{eq:hamil}
\end{equation}
defined on an $L\times N$ lattice where $l=1,2,.L$ and $n=1,2,..N$, and ${\vec
s}_{l,n}$ is a spin-1/2 operator at site $(l,n)$. The
intra-chain exchange coupling constant is $J^\prime$ and the interchain
coupling
constant is $J$, and $h$ is a magnetic field.
We use periodic boundary conditions in both directions, and as we see
below it makes a big difference.
We are interested in regime where $J^\prime < J$, and the case of odd-numbered
coupled chains, $L=3,5..$
We can think of the system as consisting of $N$ sites with $2^L$
degrees of freedom at each site, and a complicated bond interaction between the
sites mediated by the $J^\prime$ bonds in the Hamiltonian given above..
Starting with an unperturbed system with $J^\prime=0$,
we proceed to include the effect of a non-zero $J^\prime$ within a
perturbation theory,
This is achieved by identifying the most relevant degrees of freedom for a
given site, and projecting out the not so important degrees of freedom.
Below we explicit show an analytical construction of the effective-spin chain
Hamiltonian
for the case of $L=3,$ and 5, and for a larger $L$ one can a similar procedure
can be implemented numerically.

Let us consider the case of $J^\prime=0$, and $h=0$. The ground state of the
system is direct product of $N$ copies of the ground state of a closed chain
with $L$ sites.
The ground state of Heisenberg Hamiltonian of a closed chain with an odd number
of sites is in sector with
total spin $S={1\over2}$ and $S^z=\pm {1\over2}$, implying at least a  two-fold
Kramers degeneracy. As a matter of fact the ground state is four-fold
degenerate.
This extra degeneracy arises from chirality.
This can be viewed as an unpaired spin in an otherwise singlet ground state,
$i.e.$ a spinon which is forced to be there in the ground state\cite{kn:Tak81}.
In addition to the spin, the spinon carries a finite ground state momentum
$\pm 4\pi/L$ (orbital angular momentum about the center of the closed chain).
In the case of an open chain the chiral degeneracy disappears, and we will be
left with a two-fold Kramers-degenerate ground state with $S^z=\pm 1/2$.
The two ground states in $S^z={1\over2}$ sector
can be written as $\phi_{{1\over2},\pm}^L=\sum  a^L_{\pm}(\{n_i\})|\{n_i\}>$,
where $n_i$
refer to locations of down spins ($n_1 <n_2...$). Similarly the ground states
in $S^z=-{1\over2}$ sector cab be written (here $n_i$ refer to the up spins).
The amplitudes $a^L_{\pm}$ are found using Bethe anstaz\cite{kn:Gri64}; for
$L=3$ the amplitudes are given by (there is only one down spin in this case)
$a^3_{\pm}(n_1)=exp(\pm i2\pi n_1/3)$, and for $L=5$ the amplitudes are
given by
(there are two down spins in this case)
\begin{equation}
a^5_{\pm}(n_1,n_2)=\exp{\pm 2ik(n_1+n_2)} \cos((k-\eta/5)(n_2-n_1)
+\eta/2),
\label{eq:bethe}
\end{equation}
where $k=2\pi/5$ and the phase shift $\eta=2.243..$. These two different
states have spin currents going in two different directions.
We refer to these two states as having chirality by defining
$\chi^z|\phi^L_\pm>
=\pm {1\over2}|\phi^L_\pm>$, and the associated raising and lowering operators
through $\chi^\pm|\pm>=0,\chi^\pm|\mp>=|\pm>$.
The chirality operator ${\vec \chi}$ can be written in terms of the original
spin operators.
Following \cite{kn:Sub94}, we define a permutation operator
$P_l$ at a site $l$ which
permutes the spin labels such that
$l+m\rightarrow N+l-m$. This is equivalent to doing a reflection on a regular
polygon with $L$ sites, around a straightline bisecting the angle at site $l$.
The $P$ operators can be readily constructed from the
spin operators, as each of them involve $(L-1)/2$ pair-wise permutations, as
$P_l=2^{N-1\over2} \prod_{m=1}^{N-1\over2} (s_{l+m}.s_{N+l-m}+{1\over4})$.
Also, $P_l$ can be expressed in terms of chiral lowering and raising operators
as $P_l=A_l^* \chi^+ + A_l \chi^-$,
where $A_l=\exp(i2\pi l/L)$. The chirality operator is now related to
the original spin operators through $[P_l,P_{l+m}]=4i \chi^z\sin{2\pi m/L}$,
and $\chi^x=P_1/2$.

If we denote the gap in the spectrum for this $L-$site closed chain by
$\delta_L$ (see Table 1 for the actual values of the gap for $L=3,5,7,9$),
for $J^\prime <\delta_L$ it is a good approximation to just keep
the ground state manifold, $4^N$-fold degenerate for $J^\prime =0$, and
dropping
the excited states. This is expected to suffice for understanding the
low-energy behaviour of the system described by the Hamiltonian given in
Eq.~\ref{hamil}, as the excitation processes to include the the states dropped
have high energy. The ground state subspace we keep, belonging to spin-1/2
sector of each site (of $2^L$ states), is identified as the most relevant
degrees of freedom. For $L=3$, this procedure keeps all the spin-1/2 at each
site. However, for larger $L>3$, there are also spin-1/2 excited states with
large energy which are eliminated in our procedure.
This approximation is not so good for $L=9$ or more, as the
the range of $J^\prime$ for validity of our procedure becomes very restricted,
$viz. ~J^\prime<0.8612$ for $L=9$. The ground-state subspace for $J^\prime=0$,
is only $2^N$
dimensional if we use open boundary conditions for the $L-$site chain, in
contrast for periodic boundary conditions we get a much larger Hilbert space,
rendering the perturbation theory to be valid in a smaller regime of $J^\prime$
as the gap for the $L-$site open chain is smaller than that of a closed chain.
Each of the basis states ${\psi_i}$ for the
full system is a direct product of $N$ ground state eigenfunctions of close
chain of $L$ sites.
$\psi_i = \phi^1_{l_1} \times \phi^2_{l_2} ... \times \phi^{N}_{l_{N}}$,
where $l_i=1,4$.
To find the effective Hamiltonian within this subspace,
we compute the matrix elements $<{\psi_i}| H^\prime| {\psi_j}>$, and
use them to construct an effective Hamiltonian. The prime
denotes that only $J^\prime$bonds are included. The evaluation
of the matrix elements reduces to a two-site problem (actually two closed
chains of length $L$ connected by $L$ bonds of strength $J^\prime$) problem as
we have pair-wise interactions only.

Let us use two
spin-$1\over2$
effective degrees of freedom for constructing an effective Hamiltonian,
the spin and the chiral quantum numbers
${\vec S_i}$ and the chirality ${\vec \chi}_i$of the $L-$site closed chains.
An explicit calculation shows that the effective interaction strength
$\tilde A_{i,j}$ between
two closed chains connected by a $J^\prime$ bond (the corresponding operator is
${\vec s}_{i,1}.{\vec s}_{j,2}$) at sites $i$ and $j$ is
given in terms of the total spins of the two closed chains and the permutation
operators, which we discussed above,  at sites $i$ and $j$ respectively of two
chains as
\begin{equation}
{\tilde A_{i,j}}=J^\prime \alpha_L ~~{\vec S}_1 . {\vec S}_2~~ (
\gamma_L + {P_i(1)\over 2})~(\gamma_L+ {P_j(2)\over 2}).
\label{eq:effbond}
\end{equation}
Here for $L=3$,
$\alpha_3=16/9, \gamma_3=1/4$, and for $L=5$
$\alpha_5=16\beta, \gamma_5=1/20\sqrt{
\beta}$. The parameter $\beta$ is given in terms of $x_1=\cos{k-\eta/5+
\eta/2}$, and $x_2=\cos{2(k-\eta/5)+\eta/2}$ as
\begin{equation}
\beta = {2\over5} {(1+\cos{2k})x_1^4+(1+\cos{k})x_2^4-x_1^2x_2^2 \over
x_1^2+x_2^2}.
\end{equation}
For $L>5$, the wave-function renormalization constants $\alpha_L$ and
$\gamma_L$
are difficult to calculate analytically, as the number of $s^z$-basis states
in $S^Z=1/2$ becomes very large. However, the constants can be calculated
numerically, and the form of the interaction remains valid. Numerically
computed values of the constants are given in Table 1 for up to $L=9$.

The interaction $H^\prime_{12}$ between two closed chains is mediated by
$L$ bonds of strength $J^\prime$, $H^\prime_{12}=J^\prime \sum_i {\vec
s}_{i,1}.
{\vec s}_{i,2}$,
which in terms of the effective degrees of freedom
is obtained by $\sum_{i=}{\tilde A_{i,i}}$.
Now it is strightforward to write the effective Hamiltonian for the full system
within the closed chain ground state subspace as
\begin{equation}
H_{eff}=K_L\sum_{i=1}^N {\vec S_i}.{\vec S_{i+1}}(b_L +\chi^x_i\chi^x_{i+1}+
\chi^y_i\chi^y_{i+1} ),
\label{eq:effham}
\end{equation}
where $K_L=J^\prime \alpha_L L/2$, and $b_L=2 \gamma_L^2$ ($K_3=8 J^\prime/3,
b_3=1/8$, and $K_5\approx 2.845 J^\prime, b_5\approx 0.06528$, see Table 1).
It is interesting
to note that only xy-component of the chiral interaction enters the effective
Hamiltonian leaving only a two-fold up-down symmetry for the chiral variables.
The rotational invariant for the total spin is expected as we started with a
fully rotational invariant interaction and we did not break the symmetry by
restricting the Hilbert space to the spin-1/2 sector of every closed chain.
Just by looking at the form of the effective interaction we learn that on a
pair of neighbouring sites the spins and chiral variables are strongly
correlated. If we try to lower the spin energy by forcing a singlet state
between a pair of neighbouring sites, the chiral variables go into a triplet
state (with
$\chi^z=0$). However, because of the additive constant $b_L$ in front of the
chirality interaction, spin triplet and a chiral singlet is
not favored.

Let us try some variational ansatz.
A simple variational state of spin singlets and chiral triplets
on nearest neighbours, $\psi_{ST}=\prod_{i odd} \psi_i^S \psi_i^T$, where
$\psi_i^S$ is a spin singlet between sites $i$ and $i+1$ and $\psi_i^T$ is a
chiral triplet state with $\chi^z=0$, gives
an upper bound on the ground state energy of the effective Hamiltonian given
above $E_eff^G/K_L \le E_{ST}(2)/K_L=-{3\over4}(b_L+{1\over2}){N\over2}$
(=-0.9375$N/4,$ for $L=3$).
The
factor -3/4 is the spin singlet energy, one half is the chiral triplet energy
and $N/2$ is due to the fact that only half the bonds have spin singlets and
chiral triplets. This state is of course too simple to give us any insight
about
the structure of the actual ground state. However, as we will see below, it is
doing better than expected. Let us use Jordon-Wigner fermion language for the
chirality terms. The xy-term translates into a simple spinless fermion hopping.
Now it is tempting to try a free fermion ground state for chirality variables,
with an average value of ${1\over \pi}$ for the xy chiral interaction.
In this case, we would be left with a Bethe ansatz ground state for the spins,
giving an energy estimate $E_{BF}/K_L=-(\log{2} -{1\over4})(b_L+ {1\over \pi
})N (\approx -0.786 N/4$ for $L=3$). Compared to the singlet-triplet valnece
bond variational
state the Bethe Free-Fermion variational state has 20\% larger energy,
indicating that a correlated state with spin singlets and chiral triplets
with longer-ranged valence bonds would do much better. Let us consider a
four-site system and construct valence-bond state space. We form two orthogonal
states for a four-site system (the sector with $S=0$ has only two states in
this case)
from the valence bond states, $viz.$ $A=(12)(34),
B=\{(23) (41) -{1\over2}(12)(34)\}2/\sqrt{3}$, where (12) stands for a spin
singlet
between the two sites. Similarly we construct the chiral states $A^\prime$ and
$B^\prime$ where (12) denotes a chiral triplet with $\chi^z=0$. Diagonalizing
the effective Hamiltonian given in Eq.~\ref{eq:effham}, with this four-fold
spin-chiral valence bond subspace yields a variational energy of
$E_{ST}(4)/K_L\approx -0.9695 N/4$ for $L=3$. This is a 3\% lowering of energy
compared to the nearest-neighbour valence bond state we considered above. The
probability amplitude for a valence bond configuration with a bond occupied
by a spin singlet and chiral triplet is more than the other configurations,
implying strong correlations between the spin and chiral degrees of freedom. It
is interesting to note that this further neighbor valence bond variational
state has an energy very close to the exact ground state energy by less than
three percent (the indication from numerical diagonalization of the effective
Hamiltonian on a finite site system up to $N=10$ is that the ground state
energy
tends to -$N K_L/4$ for $L=3$, as we discuss below).

An effective way of constructing a convergent series of upper bounds on the
exact ground state energy is by calculating the ground state energy $E(n)$ of a
finite open chain
of $n-even$ sites, giving rise a variational inequality  $E_G/N \le E(n)/n$.
The best bound is obtained from the biggest open chain that can be numerically
diagonalized. Since the number of states per site is 4, we are able diagonalize
only up to ten sites. This gives us a bound for $L=3$
\begin{equation}
E_{GS}/NK_3  \le E(10)/10 \approx -0.9932/4
\label{eq:estim}
\end{equation}
{}From this calculation it seems that the ground state energy is tending
towards
-$NK_L/4$ from above. Similarly one can construct a sequence of lower bounds
variationally from $l-$site chain energy through $E_{G}/N \ge E(n)/(n-1)$.
However, this sequence is seen to be not as convergent as the sequence of upper
bounds. A better convergent sequence of lower bounds can be constructed by
varying the bond strengths for a given $n-$site chain, which becomes very
complicated for even $n=6$, and we will not present the details here.
The ground state is seen to be in a sector with $S=0$ and $\chi^z=0$, in all
the finite chain diagonalizations we have done up to $N=10$. The spin gap,
defined as the difference between the energies of the lowest energy states
of the sector $S=1,\chi^z=0$, and $S=0,\chi^z=0$, is seen to fall of as the
number of sites is increased, as shown in Figure 1, for different values of
$L$.
However, the data at hand is insufficient to fit the finite-$N$ scaling of the
gap. The gap seems to be non-monotonic as a function of $L$ for a given value
of $N$, as it can be seen from the figure for $L=9$ the gap is larger than
for $L=3$. This could be an artifact of our approximation, implying the
perturbation theory in conjunction with the truncation we have implemented may
not be valid for larger $L$. We have studied the spin-spin correlation function
also for the ten-site chain and is seen to be slowly decaying as in the case
of $L=1$. We may conjecture that the decay is likely to be slower for a larger
$L$ than that for the
case of $L=1$, though we are not able to substantiate it just now. This is
based
on that fact that we have a long-ranged valence bond ground state, and the
probability amplitude for a configuration with a valence bond between two
sites far apart would be much smaller for a larger $L$ than for the case of
$L=1$. But the characterization of the long-distance behaviour needs an
investigation of larger chains. The density matrix renormalization approach
\cite{Whi94} can be used for this; currently an investigation
is in progress incorporating this scheme
for the finite-size spectrum and thermodynamics of larger chains ($N>10$) for
the effective Hamiltonian.

Now we consider the case of a non-zero magnetic field in the Hamiltonian given
in Eq.
\ref{eq:hamil}. It is easy to see that the effective Hamiltonian given
in Eq.~\ref{eq:effham} will have an additional piece, $viz.$ $-h\sum S^z_i$.
Let us focus on the case of $L=3$.
The effective Hamiltonian is valid in the regime $h< 3J/4$, for $L=3$, to
ensure that the ground state subspace be composed of $S^z=\pm 1/2$ states of
the $L-$site chains. Now, if the magnetic field is strong enough to polarize
the spins in the effective Hamiltonian, the ground state would be just a
polarized spins and a free fermion state for the chiral degrees of freedom.
The range of $h$ for which this state has lower energy than $S=0$ state works
out to be
for $0.75 J > h> (7\pi-8)J^\prime/6\pi \approx 0.74225 J^\prime$. The free
fermion chiral state has  low-lying excited states with a gap
of $\Delta=(7\pi -8)K_3/4N$. This should be contrasted with the case of $L=1$,
where there are no low-lying states if the magnetic
field  is of the order of the exchange coupling constants.

In conclusion, we have established an effective spin-chiral chain Hamiltonian
for a system of odd-numbered coupled Heisenberg spin chains.
The effective interaction has a product form, $viz.$ isotropic Heisenberg for
spins and only xy interaction for chiral spins. The spin gap studied
numerically using
the effective Hamiltonian for $L=3,5,7,9$ falls off as the number of sites i
increases. Indications from a finite-size
study are that long-ranged resonating valence bond variational states (with
spin singlets and chiral triplets) are
suited to study the behaviour of the effective Hamiltonian.

It is a pleasure to thank A. M. Sengupta and M. Barma for discussions.

%\end{document}
%%-self numbering sections-%%

\vskip .2in

\newpage
\begin{tabular}{|c|c|c|c|}
\hline
$L$&$\delta_L/J$&$b_L$&$K_L/J^\prime$\\
\hline
3&3/2&1/8&8/3\\
\hline
5&1.1180&0.0697&2.8685\\
\hline
7&1.0489&0.0436&3.2765\\
\hline
9&0.8612&0.0230&4.8275\\
\hline
\end{tabular}
\vskip 3 cm
\noindent{\bf Table 1:}The energy gap of a $L-$site ring, and the wave-function
renormalization constants that appear in the effective Hamiltonian given in
Eq.~\ref{eq:effham} for various values of $L$.
\vskip 5 cm
\centerline{\bf Figure Caption}
\vskip 1cm
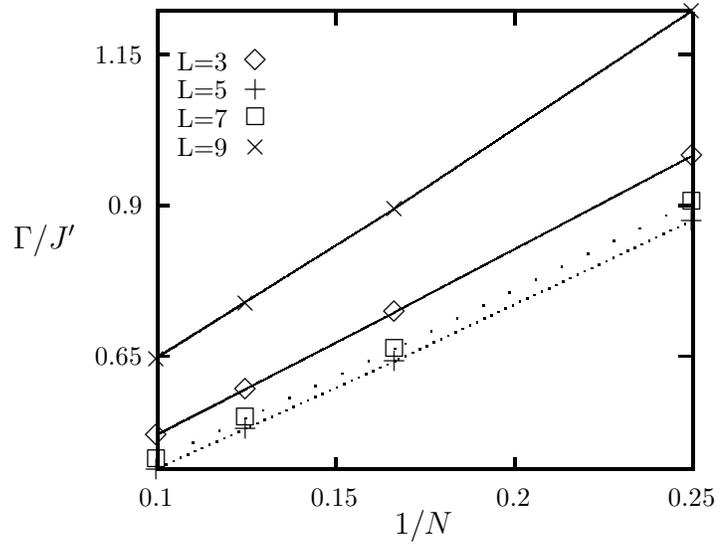
\noindent{\bf Figure 1.} The spin gap of an $L\times N$ system as a function of
$1/N$ for
various number of coupled chains, $L=3,5,7,9$. The gap is calculated
numerically from the effective Hamiltonian given in $Eq.~\ref{eq:effham}$.

\newpage
% GNUPLOT: LaTeX picture
\setlength{\unitlength}{0.240900pt}
\ifx\plotpoint\undefined\newsavebox{\plotpoint}\fi
\sbox{\plotpoint}{\rule[-0.500pt]{1.000pt}{1.000pt}}%
\begin{picture}(1125,900)(0,0)
\font\gnuplot=cmr10 at 10pt
\gnuplot
\sbox{\plotpoint}{\rule[-0.500pt]{1.000pt}{1.000pt}}%
\put(220.0,290.0){\rule[-0.500pt]{4.818pt}{1.000pt}}
\put(198,290){\makebox(0,0)[r]{0.65}}
\put(1041.0,290.0){\rule[-0.500pt]{4.818pt}{1.000pt}}
\put(220.0,527.0){\rule[-0.500pt]{4.818pt}{1.000pt}}
\put(198,527){\makebox(0,0)[r]{0.9}}
\put(1041.0,527.0){\rule[-0.500pt]{4.818pt}{1.000pt}}
\put(220.0,764.0){\rule[-0.500pt]{4.818pt}{1.000pt}}
\put(198,764){\makebox(0,0)[r]{1.15}}
\put(1041.0,764.0){\rule[-0.500pt]{4.818pt}{1.000pt}}
\put(220.0,113.0){\rule[-0.500pt]{1.000pt}{4.818pt}}
\put(220,68){\makebox(0,0){0.1}}
\put(220.0,812.0){\rule[-0.500pt]{1.000pt}{4.818pt}}
\put(500.0,113.0){\rule[-0.500pt]{1.000pt}{4.818pt}}
\put(500,68){\makebox(0,0){0.15}}
\put(500.0,812.0){\rule[-0.500pt]{1.000pt}{4.818pt}}
\put(781.0,113.0){\rule[-0.500pt]{1.000pt}{4.818pt}}
\put(781,68){\makebox(0,0){0.2}}
\put(781.0,812.0){\rule[-0.500pt]{1.000pt}{4.818pt}}
\put(1061.0,113.0){\rule[-0.500pt]{1.000pt}{4.818pt}}
\put(1061,68){\makebox(0,0){0.25}}
\put(1061.0,812.0){\rule[-0.500pt]{1.000pt}{4.818pt}}
\put(220.0,113.0){\rule[-0.500pt]{202.597pt}{1.000pt}}
\put(1061.0,113.0){\rule[-0.500pt]{1.000pt}{173.207pt}}
\put(220.0,832.0){\rule[-0.500pt]{202.597pt}{1.000pt}}
\put(45,472){\makebox(0,0){$\Gamma /J^\prime $}}
\put(640,23){\makebox(0,0){$1/N$}}
\put(640,877){\makebox(0,0){Figure 1}}
\put(220.0,113.0){\rule[-0.500pt]{1.000pt}{173.207pt}}
\put(332,754){\makebox(0,0)[r]{L=3}}
\put(376,754){\raisebox{-.8pt}{\makebox(0,0){$\Diamond$}}}
\put(1061,604){\raisebox{-.8pt}{\makebox(0,0){$\Diamond$}}}
\put(594,358){\raisebox{-.8pt}{\makebox(0,0){$\Diamond$}}}
\put(360,237){\raisebox{-.8pt}{\makebox(0,0){$\Diamond$}}}
\put(220,165){\raisebox{-.8pt}{\makebox(0,0){$\Diamond$}}}
\sbox{\plotpoint}{\rule[-0.175pt]{0.350pt}{0.350pt}}%
\put(1061,604){\usebox{\plotpoint}}
\multiput(1057.88,603.02)(-0.951,-0.500){489}{\rule{0.752pt}{0.120pt}}
\multiput(1059.44,603.27)(-465.439,-246.000){2}{\rule{0.376pt}{0.350pt}}
\multiput(590.83,357.02)(-0.970,-0.500){239}{\rule{0.764pt}{0.120pt}}
\multiput(592.41,357.27)(-232.414,-121.000){2}{\rule{0.382pt}{0.350pt}}
\multiput(356.81,236.02)(-0.977,-0.500){141}{\rule{0.768pt}{0.121pt}}
\multiput(358.41,236.27)(-138.406,-72.000){2}{\rule{0.384pt}{0.350pt}}
\put(220,165){\usebox{\plotpoint}}
\sbox{\plotpoint}{\rule[-0.300pt]{0.600pt}{0.600pt}}%
\put(332,709){\makebox(0,0)[r]{L=5}}
\put(376,709){\makebox(0,0){$+$}}
\put(1061,503){\makebox(0,0){$+$}}
\put(594,282){\makebox(0,0){$+$}}
\put(360,176){\makebox(0,0){$+$}}
\put(220,113){\makebox(0,0){$+$}}
\sbox{\plotpoint}{\rule[-0.250pt]{0.500pt}{0.500pt}}%
\put(1061,503){\usebox{\plotpoint}}
\multiput(1061,503)(-11.256,-5.327){42}{\usebox{\plotpoint}}
\multiput(594,282)(-11.344,-5.139){21}{\usebox{\plotpoint}}
\multiput(360,176)(-11.356,-5.110){12}{\usebox{\plotpoint}}
\put(220,113){\usebox{\plotpoint}}
\put(332,664){\makebox(0,0)[r]{L=7}}
\put(376,664){\raisebox{-.8pt}{\makebox(0,0){$\Box$}}}
\put(1061,531){\raisebox{-.8pt}{\makebox(0,0){$\Box$}}}
\put(594,301){\raisebox{-.8pt}{\makebox(0,0){$\Box$}}}
\put(360,192){\raisebox{-.8pt}{\makebox(0,0){$\Box$}}}
\put(220,127){\raisebox{-.8pt}{\makebox(0,0){$\Box$}}}
\put(1061,531){\usebox{\plotpoint}}
\multiput(1061,531)(-40.963,-20.175){12}{\usebox{\plotpoint}}
\multiput(594,301)(-41.392,-19.281){6}{\usebox{\plotpoint}}
\multiput(360,192)(-41.416,-19.229){3}{\usebox{\plotpoint}}
\put(220,127){\usebox{\plotpoint}}
\sbox{\plotpoint}{\rule[-0.500pt]{1.000pt}{1.000pt}}%
\put(332,619){\makebox(0,0)[r]{L=9}}
\put(376,619){\makebox(0,0){$\times$}}
\put(1061,832){\makebox(0,0){$\times$}}
\put(594,521){\makebox(0,0){$\times$}}
\put(360,374){\makebox(0,0){$\times$}}
\put(220,286){\makebox(0,0){$\times$}}
\sbox{\plotpoint}{\rule[-0.175pt]{0.350pt}{0.350pt}}%
\put(1061,832){\usebox{\plotpoint}}
\multiput(1058.46,831.02)(-0.752,-0.500){619}{\rule{0.613pt}{0.120pt}}
\multiput(1059.73,831.27)(-465.728,-311.000){2}{\rule{0.307pt}{0.350pt}}
\multiput(591.32,520.02)(-0.798,-0.500){291}{\rule{0.645pt}{0.120pt}}
\multiput(592.66,520.27)(-232.662,-147.000){2}{\rule{0.322pt}{0.350pt}}
\multiput(357.33,373.02)(-0.798,-0.500){173}{\rule{0.644pt}{0.121pt}}
\multiput(358.66,373.27)(-138.663,-88.000){2}{\rule{0.322pt}{0.350pt}}
\put(220,286){\usebox{\plotpoint}}
\end{picture}
\end{document}